\documentclass[12pt,a4paper]{article}

\usepackage{amsmath,amssymb}
\usepackage[nosort]{cite}
\usepackage[hyperref,bulletsep]{collect}
\usepackage{color}

\definecolor{grey}{rgb}{0.5,0.5,0.5}

\makeatletter \@addtoreset{equation}{section} \makeatother

\newenvironment{enum}[1][$\bullet$]{\begin{list}{#1}{\leftmargin\parindent\itemsep3pt\parsep2pt\topsep5pt}}{\end{list}}

\makeatletter
\let\old@startsection=\@startsection
\let\oldl@section=\l@section
\renewcommand{\@startsection}[6]{\old@startsection{#1}{#2}{#3}{#4}{#5}{#6\mathversion{bold}}}
\renewcommand{\l@section}[2]{%
\vspace{-0.5em}%
\oldl@section{\mathversion{bold}#1}{#2}}
\makeatother

\makeatletter
\let\old@makecaption=\@makecaption
\def\@makecaption{\small\old@makecaption}
\makeatother

\let\oldPhi=\Phi
\let\oldPsi=\Psi
\let\oldGamma=\Gamma
\let\oldDelta=\Delta
\let\oldSigma=\Sigma
\let\oldTheta=\Theta
\let\oldPi=\Pi
\renewcommand{\Phi}{\mathnormal{\oldPhi}}
\renewcommand{\Psi}{\mathnormal{\oldPsi}}
\renewcommand{\Gamma}{\mathnormal{\oldGamma}}
\renewcommand{\Sigma}{\mathnormal{\oldSigma}}
\renewcommand{\Delta}{\mathnormal{\oldDelta}}
\renewcommand{\Theta}{\mathnormal{\oldTheta}}
\renewcommand{\Pi}{\mathnormal{\oldPi}}

\newcommand{\superN}{\mathcal{N}}

\newcommand{\ham}{\mathcal{H}}
\newcommand{\dil}{\mathcal{D}}
\newcommand{\gym}{g\indups{YM}}

\newcommand{\Tr}{\mathop{\mathrm{Tr}}}

\newcommand{\order}[1]{\mathcal{O}(#1)}

\newcommand{\hateq}{\mathrel{\widehat=}}

\newcommand{\Reals}{\mathbb{R}}
\newcommand{\Sphere}{S}
\newcommand{\AdS}{AdS}

\ifx\genfrac\sdflkaj
\newcommand{\atopfrac}[2]{{{#1}\above0pt{#2}}}
\else
\newcommand{\atopfrac}[2]{\genfrac{}{}{0pt}{}{#1}{#2}}
\fi
\newcommand{\sfrac}[2]{{\textstyle\frac{#1}{#2}}}

\newcommand{\ihalf}{\sfrac{i}{2}}

\newcommand{\indups}[1]{_{\mathrm{\scriptscriptstyle #1}}}

\newcommand{\alg}[1]{\mathfrak{#1}}
\newcommand{\grp}[1]{\mathrm{#1}}

\newcommand{\lrbrk}[1]{\left(#1\right)}
\newcommand{\bigbrk}[1]{\bigl(#1\bigr)}

\newcommand{\bigcomm}[2]{\big[#1,#2\big]}
\newcommand{\comm}[2]{[#1,#2]}

\newcommand{\acomm}[2]{\{#1,#2\}}

\newcommand{\abs}[1]{{|#1|}}

\newcommand{\PTerm}[1]{\{#1\}}
\newcommand{\set}[1]{\{#1\}}
\newcommand{\state}[1]{\mathopen{|}#1\mathclose{\rangle}}

\newcommand{\nn}{\nonumber}
\newcommand{\nln}{\nonumber\\} 
\newcommand{\nl}[1][0pt]{\nonumber\\[#1]&\hspace{-4\arraycolsep}&\mathord{}}

\newcommand{\earel}[1]{\mathrel{}&\hspace{-2\arraycolsep}#1\hspace{-2\arraycolsep}&\mathrel{}}
\newcommand{\eq}{\earel{=}}

\def\[{\begin{equation}}
\def\]{\end{equation}}
\def\<{\begin{eqnarray}}
\def\>{\end{eqnarray}}

\makeatletter
\def\mr@ignsp#1 {\ifx\:#1\@empty\else #1\expandafter\mr@ignsp\fi}%
\newcommand{\multiref}[1]{\begingroup
\xdef\mr@no@sparg{\expandafter\mr@ignsp#1 \: }%
\def\mr@comma{}%
\@for\mr@refs:=\mr@no@sparg\do{\mr@comma\def\mr@comma{,}\ref{\mr@refs}}%
\endgroup}
\makeatother

\newcommand{\hypref}[2]{\ifx\href\asklfhas #2\else\href{#1}{#2}\fi}

\newcommand{\secref}[1]{Sec.~\multiref{#1}}

\newcommand{\appref}[1]{App.~\multiref{#1}}

\newcommand{\tabref}[1]{Tab.~\multiref{#1}}

\newcommand{\figref}[1]{Fig.~\multiref{#1}}
\renewcommand{\eqref}[1]{(\multiref{#1})}

\ifx\href\asklfhas\newcommand{\href}[2]{#2}\fi
\newcommand{\arxivno}[1]{\href{http://arxiv.org/abs/#1}{#1}}

\newcommand{\Q}{\mathcal{Q}}      
\newcommand{\A}{\mathcal{A}}      
\newcommand{\calL}{\mathcal{L}}   
\newcommand{\calP}{\mathcal{P}}   

\newcommand{\Md}{M^\dag}

\newcommand{\Zd}{Z^\dag}
\newcommand{\Wd}{W^\dag}
\newcommand{\bYd}{\bar Y^\dag}

\newcommand{\dynkin}[1]{[#1]}

\newcommand{\mathline}{-\!-}

\newcommand{\unit}{\mathbf{1}}

\begin{document}
\setcounter{page}{0}

\thispagestyle{empty}
\begin{flushright}\footnotesize
\texttt{\arxivno{hep-th/0510124}}\\
\texttt{AEI-2005-157}\\
\texttt{PUTP-2178}\\
\texttt{UUITP-19/05}\\
\vspace{0.5cm}
\end{flushright}
\vspace{0.5cm}

\renewcommand{\thefootnote}{\fnsymbol{footnote}}
\setcounter{footnote}{0}

\begin{center}
{\Large\textbf{\mathversion{bold}%
Long-range $\alg{gl}(n)$ Integrable Spin Chains\\
and Plane-Wave Matrix Theory}\par}
\vspace{1cm}

\textsc{N.~Beisert$^a$ and T.~Klose$^{b,c}$}
\vspace{5mm}

\textit{$^{a}$ Joseph Henry Laboratories\\
Princeton University\\
Princeton, NJ 08544, USA} \vspace{3mm}

\textit{$^{b}$ Department of Theoretical Physics, Uppsala University\\
P.O.~Box 803, SE-75108, Uppsala, Sweden}\vspace{3mm}

\textit{$^{c}$ Max-Planck-Institut f\"ur Gravitationsphysik\\
Albert-Einstein-Institut\\
Am M\"uhlenberg 1, D-14476 Golm, Germany}\vspace{3mm}

\texttt{nbeisert@princeton.edu}\\
\texttt{thomas.klose@teorfys.uu.se}\\
\par\vspace{1cm}

\vfill

\textbf{Abstract}\vspace{5mm}

\begin{minipage}{12.7cm}
Quantum spin chains arise naturally 
from perturbative large-$N$ field theories and matrix models.
The Hamiltonian of such a model is a long-range deformation 
of nearest-neighbor type interactions.
Here, we study the most general long-range integrable spin chain with
spins transforming in the fundamental representation of $\alg{gl}(n)$.
We derive the Hamiltonian and the corresponding asymptotic Bethe 
ansatz at the leading four perturbative orders
with several free parameters. 
Furthermore, we propose Bethe equations 
for all orders and identify the moduli of the integrable system.
We finally apply our results to plane-wave matrix theory and
show that the Hamiltonian in a closed 
sector is not of this form
and therefore not integrable beyond 
the first perturbative order.
This also implies that the complete model is \emph{not} integrable.
\end{minipage}

\vspace*{\fill}

\end{center}

\newpage
\setcounter{page}{1}
\renewcommand{\thefootnote}{\arabic{footnote}}
\setcounter{footnote}{0}

\tableofcontents
\vspace{10mm}
\hrule
\vspace{10mm}

\section{Introduction}

The discovery of integrability in the AdS/CFT correspondence \cite{Minahan:2002ve}
has raised interest in types of integrable spin chain models 
which were hitherto considered as somewhat exotic.
These do not only involve the usual nearest-neighbor interactions 
of two spins, but also interactions of a larger number of nearby spins
\cite{Beisert:2003tq,Beisert:2003ys}.
Although the construction of such a system appeared 
to be virtually impossible when demanding exact integrability, 
several interesting models have been found which display
convincing signs of \emph{perturbative integrability} \cite{Beisert:2003tq}:
In these models one starts with an exactly integrable nearest-neighbor
Hamiltonian and deforms it with local interactions of a longer range.
The integrable structure is only required to close up to
the considered perturbative order in the deformation parameter.
However, it should in principle be possible to extend the analysis to an arbitrary order.

In these systems, the range of the interactions grows linearly 
with the perturbative order.
Hence, if one were to transcend the
perturbative regime, one would find a truly \emph{long-ranged} spin chain. 
The only examples of such integrable spin chains 
where the non-perturbative part is well-understood
are the trigonometric Haldane-Shastry chain \cite{Haldane:1988gg,SriramShastry:1988gh}
and, as its hyperbolic and elliptic generalization, 
the Inozemtsev chain \cite{Inozemtsev:1989yq,Inozemtsev:2002vb}.
These long-range models involve only spin-spin interactions, but the spin chain Hamiltonians
which are relevant in the context of the
AdS/CFT correspondence appear to be more general.

In a conformal gauge theory, predominantly $\superN=4$ supersymmetric
Yang-Mills theory (SYM), these spin chains arise from the eigenvalue problem 
of the dilatation operator in the planar limit.
The interactions of growing range correspond to higher and higher
perturbative corrections to the dilatation operator at small coupling.
This relationship implies that the energy spectrum of the spin chain system
translates into the spectrum of anomalous dimensions in gauge theory.
For an integrable system one can then make use of the integrable structure 
of the spin chain system, in particular the Bethe ansatz, 
to determine planar anomalous dimensions in field theory
\cite{Minahan:2002ve,Beisert:2003yb,Serban:2004jf,Beisert:2004hm,Staudacher:2004tk,Beisert:2005fw}. 
This vastly simplifies the determination of these quantities as compared
to direct (higher-loop) computations.

Integrable structures have a long history in string theory, too. 
Of particular interest here is IIB string theory on $\AdS_5 \times \Sphere^5$ 
which is dual to $\superN=4$ gauge theory
and whose integrability was established only recently
following \cite{Mandal:2002fs,Bena:2003wd,Arutyunov:2003uj,Kazakov:2004qf}.
Intriguingly, the spectrum of this quantum string theory 
can be described by Bethe equations as well.
The equations found in \cite{Arutyunov:2004vx}
are surprisingly similar to the ones found in gauge theory \cite{Beisert:2004hm}. 
They reproduce accurately results of various direct computations 
in string theory in various cases 
\cite{Gubser:2002tv,Callan:2003xr,McLoughlin:2004dh,Frolov:2004bh,Park:2005ji,Schafer-Nameki:2005is,Beisert:2005cw}
as shown in 
\cite{Arutyunov:2004vx,Staudacher:2004tk,Beisert:2005mq,Hernandez:2005nf,Schafer-Nameki:2005tn,Beisert:2005cw}.
In fact, the stringy Bethe ansatz also has 
a spin chain realization \cite{Beisert:2004jw},
even if its regime of validity does not coincide with perturbative 
quantum strings.

In order to gain further insights into the integrable properties of gauge field theories,
closely related matrix models were studied as toy models for integrability in large-$N$ theories.
In these cases, it is the planar part of the matrix model Hamiltonian
which is interpreted as the Hamiltonian of a spin chain system.
Due to the finite number of degrees of freedom,
matrix models are technically easier to handle 
in explicit calculations than field theories.
For these models one can specifically construct an integrable Hamiltonian \cite{Lee:1997nt,Lee:1997dd}. 
Even more, in some models integrability originates 
from an underlying action principle.
A particularly interesting matrix model akin to $\superN=4$ SYM
is given by plane-wave matrix theory (PWMT) \cite{Berenstein:2002jq}.
In fact, this theory acts as the microscopic definition of M-theory
on an eleven-dimensional plane-wave \cite{Berenstein:2002jq,Dasgupta:2002hx}.
However, it was also shown to be derivable from SYM
as a consistent truncation on $\Reals \times \Sphere^3$ \cite{Kim:2003rz}.
Through this connection,
the one-loop integrability of PWMT is immediately inherited from SYM \cite{Kim:2003rz},
but also higher-loop studies uncovered integrable sectors of PWMT \cite{Klose:2003qc,Fischbacher:2004iu}.

For in-depth reviews of the above mentioned topics we would like to 
refer the reader to 
\cite{Beisert:2004ry,Beisert:2004yq,Zarembo:2004hp} 
for gauge theory, to \cite{Klose:2005aa} for matrix models
and to \cite{Tseytlin:2003ii,Plefka:2005bk} for spinning strings.
\bigskip

Currently there is a small zoo of perturbatively integrable spin chains available
which differ in one or another respect. 
These, however, might appear to be just a few known islands 
in otherwise uncharted territory. 
An underlying general framework or classification,
such as the R-matrix formalism 
or Bethe ans\"atze for generic algebras and representations,
is yet unknown, although some patterns begin to surface. 
Here we make the first attempt at a systematic approach:
We will study the most general perturbatively integrable spin chain 
with spins in the fundamental representation of $\alg{gl}(n)$ 
and long-range interactions.
We construct two fundamental spin chain operators, 
the Hamiltonian $\Q_2$ and one higher charge $\Q_3$ from the integrable structure,
without reference to some underlying large $N$ gauge theory. 
As a first approximation to full perturbative integrability,
we merely demand commutation of these two operators
\[\label{eq:PertInt}
\bigcomm{\Q_2(\lambda)}{\Q_3(\lambda)}=\order{\lambda^{k+1}} \; .
\]
Here $\lambda$ is a formally small parameter and $k$ the maximum 
perturbative order we shall be interested in.
This is a minimum requirement for integrability, 
but apparently it is also sufficient for our model (to the given accuracy).
In this article, we obtain the most general spin chain of this kind 
up to fourth order, i.e.~$k=4$ in \eqref{eq:PertInt}
(\secref{sec:spin-chain} and \appref{sec:perms}).
For the resulting Hamiltonian we perform the asymptotic
coordinate space Bethe ansatz 
\cite{Bethe:1931hc,Sutherland:1978aa,Staudacher:2004tk}
and obtain asymptotic Bethe equations
for the model (\secref{sec:Bethe}). 
This allows us to find patterns in the set of free parameters
of the model which might generalize to arbitrary perturbative orders.
For different values of these moduli we recover
various Bethe equations found in the context of AdS/CFT 
(\secref{sec:coeffs}).
\medskip

In the second half of these notes, we focus our attention back to gauge
theories and matrix models which have been the starting point for these
kinds of perturbatively integrable models. 
Our main observation is that for $\alg{gl}(n\ge3)$ 
the obtained spin chain operators at subleading orders
cannot arise from a large-$N$ gauge theory 
with at most four-valent interaction vertices
(\secref{sec:gauge-theory}). 
Conversely, as the determined spin chain is the most general integrable one,
any spin chain Hamiltonian that is derived from a large-$N$ gauge theory with at most 
four-valent interaction vertices and fields in the fundamental representation 
of $\alg{gl}(n\ge3)$ cannot be integrable beyond first order. 
This result has a severe consequence for PWMT.
When determining all closed subsectors of this theory,
we indeed find a closed fundamental $\alg{su}(3)$ sector (\secref{sec:pwmt}),
which is non-integrable by this general argument.
Hence, we have shown that PWMT is---despite the integrable properties
within some sectors at low perturbative orders---certainly not fully integrable!
In contrast, the analog fundamental $\alg{su}(3)$ sector
of $\superN=4$ SYM is not closed beyond leading order
and thus there is no obstruction to integrability.
The difference between both models is that in
the full gauge theory the corresponding fields mix with fermions;
these fermions have been projected out in PWMT,
along with, sadly, integrability.

\section{Perturbatively integrable $\alg{gl}(n)$ spin chain} 
\label{sec:spin-chain}

In the following we focus on spin chains with spins transforming in 
the fundamental representation of $\alg{gl}(n)$.%
\footnote{We expect the analysis to be equivalent if the
symmetry algebra is the superalgebra $\alg{gl}(n|m)$. 
For simplicity of notation we shall nevertheless restrict to $\alg{gl}(n)$.}
The integrability of a spin chain system 
is expressed through the existence 
of an infinite set of 
hermitian spin chain charges $\Q_r$ with $r\in\{2,3,\ldots\}$, 
all commuting with each other
\[\label{eqn:commutation}
  \comm{\Q_r}{\Q_s} = 0 \; .
\]
Here $\Q_2$ represents the spin chain Hamiltonian 
and $\Q_{r\ge3}$ are called higher charges. 
It is reasonable to define a total momentum operator
$\Q_1$ such that $\exp(i\Q_1)$ shifts all spins by one lattice site.
Note however that only the operator $\exp(i\Q_1)$ is well-defined while its
logarithm $\Q_1$ is ambiguous.

We also introduce the operator $\calL$ 
which counts the number of spins within a spin chain state, 
i.e.~which measures the spin chain length $L$. 
Since in the models under consideration $\calL$ commutes with all charges, 
the full space of states is divided into superselection sectors 
characterized by the spin chain length. 
Such a sector contains a finite number of $n^L$ states. 
Therefore the Hamiltonian $\Q_2$, restricted to a specific sector, 
is nothing but a 
$n^L \times n^L$ hermitian matrix.
It can always be diagonalized by a similarity transformation 
and the existence of $n^L-1$ further mutually commuting matrices is trivial. 
The distinguishing feature of (most) integrable models is that there
exist $L$ \emph{local} commuting charges $\Q_r$
which take a generic form independently of the value of $L$.
The notion of locality will be rendered more precisely in the following. 
Furthermore, the charges should also be invariant under the symmetry group
of the model.

The construction principle for \emph{perturbatively} integrable spin chain systems 
is the requirement that the spin chain charges 
can be expanded according to their interaction range. 
We write   
\[ \label{eqn:charge-expansion}
\Q_r(\lambda) = \sum_{k=0}^\infty \lambda^k \Q_r^{(k)} \; ,
\]
where by definition the maximal range of $\Q_r^{(k)}$ is $r+k$, 
i.e.~$\Q_r^{(k)}$ acts locally on $r+k$ adjacent spins. 
The constant $\lambda$ is used as a formal expansion parameter. 
With regard to gauge theory, 
$\lambda$ should be viewed as ('t Hooft) coupling constant 
and \eqref{eqn:charge-expansion} as a perturbation expansion.
If the perturbation expansion is truncated at some order in $\lambda$,
then the commutation \eqref{eqn:commutation} is no longer exact,
except for the case $\lambda=0$.
Therefore the starting point of the perturbative expansion is
an exactly integrable spin chain with charges $\Q_r^{(0)}$. 
It is easy to convince oneself that this is 
a common nearest-neighbor spin chain. 
Conversely, the charges $\Q_r(\lambda)$ only commute exactly
when \emph{all} terms $\Q_r^{(k)}$ are taken into account. 
The range of the charges becomes formally infinite at finite $\lambda$
and the spin chain should thus be considered a \emph{long-range} system
\cite{Beisert:2003tq}.

Note that the above definition of charges assumes
a spin chain of infinite length.
When the range $r+k$ of the interaction 
exceeds the length of a periodic state of length $L$,
we need to specify alternative rules for the application of $\Q_r$.
Unfortunately, the notion of locality breaks down at this
point because the interaction already extends over
the whole state. Therefore our main construction principle disappears
and any invariant, commuting deformation of the charges 
might be considered integrable.
In order to avoid this complication, 
we restrict the validity of $\Q_r$ 
to $\order{\lambda^{L-r}}$.%
\footnote{In gauge theories and matrix models,
there are so-called \emph{wrapping} interactions \cite{Beisert:2003ys}
which set in at this order. 
If the model contains \emph{double-trace} vertices, 
cf.~\cite{Dymarsky:2005uh,Dymarsky:2005nc},
we should further restrict the validity to 
$\order{\lambda^{L-r-1}}$ due to 
interactions involving these vertices 
\cite{Freedman:2005cg,Penati:2005hp,Rossi:2005mr}.}
Put differently, when considering the system up to order $\order{\lambda^k}$,
then we assume the spin chains to have length $L \ge k+r$ 
and call this the \emph{asymptotic} regime.

Now let us introduce a basis for the operators $\Q_r$.
We will use products of nearest-neighbor permutations 
$\calP_{i,i+1}$ to represent $\alg{gl}(n)$-invariant 
interactions of the spins \cite{Beisert:2003tq}.%
\footnote{Any $\alg{gl}(n)$-invariant endomorphism of 
a tensor product of fundamental modules can be written 
as a sum of permutations of the modules. 
Furthermore, any permutation can be decomposed
as a product of permutations of nearest neighbors.}$^,$%
\footnote{For the superalgebra $\alg{gl}(n|m)$ 
we should replace ordinary permutations by graded ones.
Note, however, that this does not include
any of the supersymmetric sectors of matrix models
or spin chains, cf.~\secref{sec:pwmt}:
There the states do not transform
in tensor products of the fundamental representations
away from $\lambda=0$;
the representation depends on $\lambda$.}
We abbreviate a sequence of these permutations as
\[ \label{eqn:permutations}
\PTerm{a_1,a_2,\ldots,a_l} 
:= \sum_{i=1}^{\calL} 
\calP_{i+a_1,i+a_1+1} \calP_{i+a_2,i+a_2+1} \cdots \calP_{i+a_l,i+a_l+1} \; , 
\qquad
\PTerm{} := \calL \; .
\]
For details and some useful identities we refer to \appref{sec:perms}. 
The range of a permutation \eqref{eqn:permutations} 
is given by $R = \max\set{a_i} - \min\set{a_i} + 2$. 
We list the permutations of a small range in \tabref{tab:permutation-operators}.
\begin{table}[t]
\begin{center}
\begin{tabular}{|l|l|} \hline
  $R$ & Permutations \\ \hline
  $1$ & $\PTerm{}$ \\ \hline
  $2$ & $\PTerm{0}$ \\ \hline
  $3$ & $\PTerm{0,1}, \PTerm{1,0} \textcolor{grey}{, \PTerm{0,1,0}}$ \\ \hline
  $4$ & $\PTerm{0,2}, \PTerm{0,1,2}, \PTerm{0,2,1}, \PTerm{1,0,2},
         \PTerm{2,1,0} \textcolor{grey}{, \PTerm{0,1,0,2}, \PTerm{0,1,2,1}, \PTerm{0,2,1,0},}$ \\
      & $\textcolor{grey}{ \PTerm{1,0,2,1}, \PTerm{1,2,1,0}, \PTerm{0,1,0,2,1},
         \PTerm{0,1,2,1,0}, \PTerm{1,0,2,1,0}, \PTerm{0,1,0,2,1,0}}$ \\ 
      & $  \#5 + \textcolor{grey}{\#9} =    \#14$ \\ \hline
  $5$ & $ \#15 + \textcolor{grey}{\#63} =   \#78$ \\ \hline
  $6$ & $ \#50 + \textcolor{grey}{\#454} =  \#504$ \\ \hline
  $7$ & $\#175 + \textcolor{grey}{\#3545} = \#3720$ \\ \hline
\end{tabular}
\end{center}
\caption{Permutation operators of range $R$. 
(The occurrence of operators printed in grey 
is restricted when the spin chain originates 
from gauge theories or matrix models 
as we will discuss in \protect\secref{sec:gauge-theory}.)}
\label{tab:permutation-operators}
\end{table}
The number of all permutations up to range $R$ is 
given by the formula $R!-(R-1)!+1$.%
\footnote{The formula is explained as follows: 
There are $R!$ local permutations with a maximum range $R$. 
However, the interactions of the form 
$\PTerm{a_1,\ldots,a_l}-\PTerm{a_1+1,\ldots,a_l+1}$
lead to telescoping sums in \eqref{eqn:permutations};
they do not act on periodic states and we discard them.
These are constructed from interactions 
of range $R-1$ except for the trivial $\PTerm{}$.
Therefore we drop $(R-1)!-1$ interactions from the original $R!$.}
The first few elements of this sequence are $1,2,5,19,97,601,4321,\ldots$\,
which agrees with our findings in \tabref{tab:permutation-operators}.%

We now construct the charges $\Q_2^{(k)}$ and $\Q_3^{(k)}$
as a linear combination of the building blocks of range $R \le r+k$
in \tabref{tab:permutation-operators}.
We also demand that the charges are hermitian, $\Q_r^{\dagger}=\Q_r$,
cf.~\eqref{eqn:permutation-operators-hermitian-conjugation}.
This leads to $(k+2)!-(k+1)!+1$ and 
$(k+3)!-(k+2)!+1$ free real coefficients for 
$\Q^{(k)}_2$ and $\Q^{(k)}_3$, respectively. 
We keep track of the number of free coefficients of the ansatz 
at a certain perturbative order in \tabref{tab:coeffs}. 
To obtain an integrable system, 
we demand that the two charges commute, 
i.e.~at order $\lambda^k$ we solve the equation
\[\label{eqn:perturbative-commutation}
  \sum_{\ell=0}^k \bigcomm{\Q_2^{(\ell)}}{\Q_3^{(k-\ell)}} = 0 \; .
\]
Taking into account the commutation with yet higher charges $\Q_r$
apparently does not lead to further restrictions for $\Q_2$ or $\Q_3$
similar to the observations in 
\cite{Grabowski:1995rb,Beisert:2003tq,Beisert:2004hm,Beisert:2004jw}.%
\footnote{For more rigorous statements, 
one might consider Yangian structures \cite{Bernard:1993ya,Dolan:2003uh,Agarwal:2004sz}.}

We have performed the computation up to and including fourth order.
Since the expressions are rather lengthy 
we present the spin chain operators 
at the end of the paper (\tabref{tab:Q2,tab:Q3}) 
and only up to second order. 
At every loop order we are left with some free parameters, 
which are not determined by \eqref{eqn:perturbative-commutation}. 
These correspond to various kinds of degrees of freedom of the system: 
The parameters $\alpha_\ell(\lambda), \beta_{r,s}(\lambda)$ are moduli 
which govern the propagation and scattering of spin flips 
induced by the application of $\Q_2$ onto the spin chain. 
These will be explained further in the next section. 
The parameters $\gamma_{r,s}(\lambda)$ correspond 
to taking linear combinations of commuting charges 
which still obey the integrability relation \eqref{eqn:commutation}. 
Finally, $\epsilon_{k,\ell}(\lambda)$ are a set of parameters 
which affect the eigenvectors of the charges, 
but not their spectrum. 
Hence they correspond to similarity transformations of the spin chain operators. 
The number of the various parameters are listed in \tabref{tab:coeffs}. 
by adding up the lines. 
Note that the number of parameters $\epsilon_{k,\ell}$ 
corresponding to similarity transformations 
matches up with our expectations: 
At order $\lambda^k$ we may use any permutation of range $R \le k$,
cf.~\tabref{tab:permutation-operators}, to generate a similarity transformation.
Note, however, that the transformations generated by 
the algebra of charges $\calL,\Q_2,\Q_3,\ldots$ are trivial
by \eqref{eqn:commutation}.
Hence there is one parameter for each permutation structure
less one for each commuting charge of suitable range.

\begin{table}
\begin{center}
\begin{tabular}{|l|c|c|c|c|c|c|} \cline{2-7}
\multicolumn{1}{l|}{} & $\pm$ & $\lambda^0$ & $\lambda^1$ & $\lambda^2$ & $\lambda^3$ & $\lambda^4$ \\ \hline
ansatz for Hamiltonian $\Q_2$                 &     & 2 & 5  & 19  & 97  & 601  \\ 
ansatz for third charge $\Q_3$                & $+$ & 5 & 19 & 97  & 601 & 4321 \\ 
constraint from commutation                   & $-$ & 2 & 16 & 102 & 666 & 4807 \\ \hline
undetermined coefficients                     & $=$ & 5 & 8  & 14  & 32  & 115  \\ 
$\alpha_{\ell}$ (rapidity map)                & $-$ & 0 & 1  & 2   &  3  & 4    \\ 
$\beta_{r,s}$ (dressing factor)               & $-$ & 0 & 0  & 1   &  3  & 6    \\
$\gamma_{2,s}$ (eigenvalue Hamiltonian)       & $-$ & 2 & 3  & 4   &  5  & 6    \\ 
$\gamma_{3,s}$ (eigenvalue third charge)      & $-$ & 3 & 4  & 5   &  6  & 7    \\ \hline 
$\epsilon_{k,l}$ (similarity transformations) & $=$ & 0 & 0  & 2   & 15  & 92   \\ 
trivial similarity transformations            & $+$ & 1 & 2  & 3   &  4  & 5    \\ \hline
all similarity transformations                & $=$ & 1 & 2  & 5   & 19  & 97   \\ \hline
\end{tabular}
\end{center}
\caption{Number of free parameters. 
The parameters $\alpha_k,\beta_{r,s}$ characterize different spin chain systems
while $\gamma_{r,s}$ fixes linear combinations of the charges.
The parameters $\epsilon_{k,l}$ correspond to similarity transformations 
of a given spin chain system and do not influence the spectrum.}
\label{tab:coeffs}
\end{table}

\section{Bethe Ansatz}
\label{sec:Bethe}

We now perform the nested asymptotic coordinate space Bethe ansatz 
\cite{Bethe:1931hc,Sutherland:1978aa}
as outlined in \cite{Staudacher:2004tk,Beisert:2005fw}.
Here, we will merely state the results of this analysis
in a form which highlights the different types of free parameters 
of the system.

A state is described by a set of Bethe roots $u_{\ell,k}$. 
The label $\ell=1,\ldots,n-1$ indicates the flavor of the Bethe root,
i.e.~its level in the nested Bethe ansatz. 
The label $k=1,\ldots,K_\ell$ indexes the set of Bethe roots
of flavor $\ell$, where $K_\ell$ is the total number of Bethe roots of that kind. 

Let us first of all present the Bethe equations and then describe 
all the various required definitions. The main Bethe equation 
at level $\ell=1$ reads
\[\label{eq:BetheMain}
1=\lrbrk{\frac{x(u_{1,k}-\ihalf)}{x(u_{1,k}+\ihalf)}}^L 
     \prod_{\textstyle\atopfrac{j=1}{j\neq k}}^{K_1}
      \lrbrk{ \frac{u_{1,k}-u_{1,j}+i}{u_{1,k}-u_{1,j}-i}\, \exp \bigbrk{2i\theta(u_{1,k},u_{1,j})} }
     \prod_{j=1}^{K_2} \frac{u_{1,k}-u_{2,j}-\ihalf}{u_{1,k}-u_{2,j}+\ihalf} \; .
\]
This equation is similar in form to the equation proposed
in \cite{Beisert:2004hm,Arutyunov:2004vx}.
The remaining auxiliary Bethe equations are 
as usual for the algebra $\alg{gl}(n)$.%
\footnote{This might also work for $\alg{gl}(n|m)$ with fundamental spins
by substituting the usual auxiliary Bethe equations for this superalgebra.}
This follows straightforwardly from the fact that the level-2 spin chain 
is of nearest-neighbor type and has manifest $\alg{gl}({n-1})$ invariance.%
\footnote{We thank M.~Staudacher for discussions of this point.}
For levels $\ell=2,\ldots,n-2$ the Bethe equations are given by
\cite{Ogievetsky:1986hu}
\[\label{eq:BetheAux}
1=\prod_{j=1}^{K_{\ell-1}} \frac{u_{\ell,k}-u_{\ell-1,j}-\ihalf}{u_{\ell,k}-u_{\ell-1,j}+\ihalf}
  \prod_{\textstyle\atopfrac{j=1}{j\neq k}}^{K_\ell}\frac{u_{\ell,k}-u_{\ell,j}+i}{u_{\ell,k}-u_{\ell,j}-i}
  \prod_{j=1}^{K_{\ell+1}} \frac{u_{\ell,k}-u_{\ell+1,j}-\ihalf}{u_{\ell,k}-u_{\ell+1,j}+\ihalf}
\]
and for the final level $\ell=n-1$ by
\[\label{eq:BetheAuxEnd}
1=\prod_{j=1}^{K_{n-2}} \frac{u_{n-1,k}-u_{n-2,j}-\ihalf}{u_{n-1,k}-u_{n-2,j}+\ihalf}
  \prod_{\textstyle\atopfrac{j=1}{j\neq k}}^{K_{n-1}}\frac{u_{n-1,k}-u_{n-1,j}+i}{u_{n-1,k}-u_{n-1,j}-i} \; .
\]
We now explain the rapidity map $x(u)$ 
and the dressing phase $\theta(u_{1,k},u_{1,j})$.

The \emph{rapidity map} $x(u)$ shall be defined implicitly through its inverse
\[\label{eq:uofx}
u(x)=x+\sum_{\ell=0}^\infty \frac{\alpha_{\ell}(\lambda)}{x^{\ell+1}} \; .
\]
This is the first place where some of the 
undetermined coefficients within the spin chain Hamiltonian 
derived in the previous section enter the Bethe ansatz. 
The freedom is such that one may choose for the $\alpha_\ell(\lambda)$ 
arbitrary series starting at order $\order{\lambda^{\ell+1}}$:
\[\label{eq:alphadef}
\alpha_{\ell}(\lambda) = \sum_{k = \ell+1}^\infty \lambda^k \alpha_{\ell}^{(k)} \; .
\]
If, however, the Hamiltonian $\Q_2$ is to conserve parity 
as defined in \eqref{eqn:permutation-operators-parity-conjugation},
then one must set $\alpha_{\ell} = 0$ for all odd $\ell$. 
The inverse map from the $u$-plane to the $x$-plane has the following form
\[\label{eq:xofu}
x(u)=\frac{u}{2}+\frac{u}{2}\sqrt{1-4\sum_{\ell=0}^\infty \frac{\tilde\alpha_{\ell}(\lambda)}{u^{\ell+2}}} \; .
\]
The parameters $\tilde\alpha_\ell(\lambda)$ are fixed uniquely
by the components of $\alpha_k(\lambda)$ in \eqref{eq:alphadef}.
Here $\tilde\alpha_0(\lambda)$ starts at order $\order{\lambda}$ 
and $\tilde\alpha_{\ell\ge1}(\lambda)$ at $\order{\lambda^{[\ell/2]+2}}$.
The coefficients $\alpha_\ell(\lambda)$ govern the propagation 
of spin flips in the ferromagnetic vacuum.

Next we present the \emph{dressing phase}
\[\label{eq:dressing}
\theta(u_1,u_2)=
\sum_{r=2}^\infty
\sum_{s=r+1}^\infty
\beta_{r,s}(\lambda)
\bigbrk{q_r(u_1)\,q_s(u_2)-q_s(u_1)\,q_r(u_2)} \; .
\]
This is a generalization of the phase proposed 
in \cite{Arutyunov:2004vx,Klose:2005aa}
to the case of a parity-violating Hamiltonian.
It is the second place where some other free coefficients of the
spin chain Hamiltonian enter the Bethe ansatz. They are in one-to-one
correspondence with the functions $\beta_{r,s}(\lambda)$ 
which start at order $\order{\lambda^{s-1}}$:
\[\label{eq:betadef}
  \beta_{r,s}(\lambda) = \sum_{k = s-1}^\infty \lambda^k \beta_{r,s}^{(k)} \; .
\]
These functions govern the scattering of two spin flips. 
Here parity conservation demands $\beta_{r,s} = 0$ for all even $r+s$. 
For \eqref{eq:dressing} we also define the \emph{elementary magnon charges} as
\[\label{eq:magnoncharge}
q_r(u)=\frac{1}{r-1}\lrbrk{\frac{i}{x(u+\ihalf)^{r-1}}-\frac{i}{x(u-\ihalf)^{r-1}}} \; .
\]

The solutions to the above Bethe equations define 
the set of periodic eigenstates of the system. 
Finally, we have to specify the eigenvalues of the Hamiltonian and higher charges
in terms of the rapidities $u_{\ell,k}$. 
First, we consider the eigenvalue of the shift operator $\exp(i\Q_1)$ 
which is given by
\[\label{eq:shifteigen}
\exp(iQ_1)=\prod_{k=1}^{K_1}\frac{x(u_{1,k}+\ihalf)}{x(u_{1,k}-\ihalf)} \; .
\]
In particular, cyclic states obey the zero-momentum condition $\exp(iQ_1)=1$. 
%
%
The eigenvalues of the spin chain charges are determined by the formula
\[\label{eq:chargeeigen}
Q_r=\gamma_{r,0}(\lambda)\,L+\sum_{s=2}^\infty \gamma_{r,s}(\lambda)\,\bar Q_s \; , \qquad
\bar Q_s= \sum_{k=1}^{K_1} q_s(u_{1,k}).
\]
The functions $\gamma_{r,s}(\lambda)$ represent a normalization matrix
of the charges $Q_r$ in terms of the normalized charges $\bar Q_s$ and the length $L$.
Again, this freedom is reflected by undetermined coefficients in the spin chain operators
which we determined in the previous section by commutation. They allow for arbitrary functions
$\gamma_{r,s}(\lambda)$ starting at order $\order{\lambda^{\max(s-r,0)}}$:
\[\label{eq:gammadef}
  \gamma_{r,s}(\lambda) = \sum_{k = \max(s-r,0)}^\infty \lambda^k \gamma_{r,s}^{(k)} \; .
\]
Note that for a system with parity conservation we must set $\gamma_{r,s}(\lambda)=0$ whenever $r+s$ is odd.

This concludes our presentation of the Bethe ansatz. 
The system with arbitrary parameters $\alpha,\beta,\gamma$ 
describes the same spectra as the commuting Hamiltonians 
we have found in \secref{sec:spin-chain} 
up to fourth order $\order{\lambda^4}$. 
The parameters $\epsilon$ merely influence the
eigenvectors and therefore do not appear in the Bethe equations.
We believe that at higher orders 
and for higher charges the conjectured form 
of the equations and the implied number of free parameters 
remains correct although we certainly 
do not claim to have a proof for our conjecture.

\section{Constraints from matrix and gauge theories} \label{sec:gauge-theory}

Our motivation to study \emph{long-range} spin chain lies in the fact
that they naturally arise in the large-$N$ 
limit of $\grp{U}(N)$ matrix models and gauge theories 
\cite{Beisert:2003tq,Kim:2003rz}.
Single-trace states%
\footnote{In gauge theory, the term `state' 
shall refer to a gauge-invariant local operator.}
are in one-to-one correspondence with cyclic states of a 
closed quantum spin chain. A matrix or a field inside the trace corresponds
to a site of the spin chain; its flavor becomes the orientation of the spin.
In the large-$N$ limit, the sector of single-trace states is closed 
and symmetry generators acting on states can be identified 
with spin chain operators.
In particular, the Hamiltonian of matrix quantum mechanics becomes
the spin chain Hamiltonian. Likewise, for a conformal gauge theory 
we can identify the dilatation generator with the spin chain Hamiltonian.%
\footnote{Here the logarithm of the radial coordinate can be
viewed as a time coordinate. In this picture the gauge theory can be viewed
as a matrix quantum mechanics of infinitely many matrices
which represent the modes of the fields in the transverse space.}
For a review of this correspondence, see \cite{Beisert:2004ry,Plefka:2005bk,Klose:2005aa}.

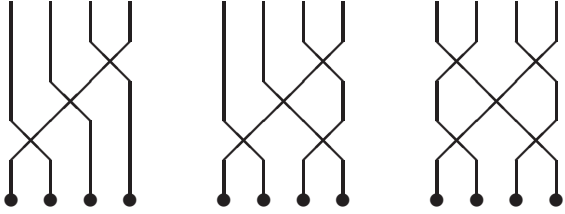
\begin{figure}\centering
\setlength{\unitlength}{1pt}%
\small\thicklines%
\begin{picture}(55,85)(-5,-5)
\put(  0, 0){\circle*{5}}%
\put( 15, 0){\circle*{5}}%
\put( 30, 0){\circle*{5}}%
\put( 45, 0){\circle*{5}}%
\put(  0, 0){\line(0,1){15}}%
\put( 15, 0){\line(0,1){15}}%
\put( 30, 0){\line(0,1){15}}%
\put( 45, 0){\line(0,1){15}}%
\put(  0,15){\line(1,1){15}}%
\put( 15,15){\line(-1,1){15}}%
\put( 30,15){\line(0,1){15}}%
\put( 45,15){\line(0,1){15}}%
\put(  0,30){\line(0,1){15}}%
\put( 15,30){\line(1,1){15}}%
\put( 30,30){\line(-1,1){15}}%
\put( 45,30){\line(0,1){15}}%
\put(  0,45){\line(0,1){15}}%
\put( 15,45){\line(0,1){15}}%
\put( 30,45){\line(1,1){15}}%
\put( 45,45){\line(-1,1){15}}%
\put(  0,60){\line(0,1){15}}%
\put( 15,60){\line(0,1){15}}%
\put( 30,60){\line(0,1){15}}%
\put( 45,60){\line(0,1){15}}%
\end{picture}
\qquad
\begin{picture}(55,85)(-5,-5)
\put(  0, 0){\circle*{5}}%
\put( 15, 0){\circle*{5}}%
\put( 30, 0){\circle*{5}}%
\put( 45, 0){\circle*{5}}%
\put(  0, 0){\line(0,1){15}}%
\put( 15, 0){\line(0,1){15}}%
\put( 30, 0){\line(0,1){15}}%
\put( 45, 0){\line(0,1){15}}%
\put(  0,15){\line(1,1){15}}%
\put( 15,15){\line(-1,1){15}}%
\put( 30,15){\line(1,1){15}}%
\put( 45,15){\line(-1,1){15}}%
\put(  0,30){\line(0,1){15}}%
\put( 15,30){\line(1,1){15}}%
\put( 30,30){\line(-1,1){15}}%
\put( 45,30){\line(0,1){15}}%
\put(  0,45){\line(0,1){15}}%
\put( 15,45){\line(0,1){15}}%
\put( 30,45){\line(1,1){15}}%
\put( 45,45){\line(-1,1){15}}%
\put(  0,60){\line(0,1){15}}%
\put( 15,60){\line(0,1){15}}%
\put( 30,60){\line(0,1){15}}%
\put( 45,60){\line(0,1){15}}%
\end{picture}
\qquad
\begin{picture}(55,85)(-5,-5)
\put(  0, 0){\circle*{5}}%
\put( 15, 0){\circle*{5}}%
\put( 30, 0){\circle*{5}}%
\put( 45, 0){\circle*{5}}%
\put(  0, 0){\line(0,1){15}}%
\put( 15, 0){\line(0,1){15}}%
\put( 30, 0){\line(0,1){15}}%
\put( 45, 0){\line(0,1){15}}%
\put(  0,15){\line(1,1){15}}%
\put( 15,15){\line(-1,1){15}}%
\put( 30,15){\line(1,1){15}}%
\put( 45,15){\line(-1,1){15}}%
\put(  0,30){\line(0,1){15}}%
\put( 15,30){\line(1,1){15}}%
\put( 30,30){\line(-1,1){15}}%
\put( 45,30){\line(0,1){15}}%
\put(  0,45){\line(1,1){15}}%
\put( 15,45){\line(-1,1){15}}%
\put( 30,45){\line(1,1){15}}%
\put( 45,45){\line(-1,1){15}}%
\put(  0,60){\line(0,1){15}}%
\put( 15,60){\line(0,1){15}}%
\put( 30,60){\line(0,1){15}}%
\put( 45,60){\line(0,1){15}}%
\end{picture}

\caption{Three planar Feynman diagrams of range $R=4$.
The left interaction contains at least 3 four-vertices and
appear at three loops, i.e.~$\order{\lambda^2}$.
The middle one contains at least 4 four-vertices and 
can appear only at four loops, i.e.~$\order{\lambda^3}$.
The right one may (in principle) represent a
fermionic loop and could also
arise at four loops.
}
\label{fig:diaginter}
\end{figure}

The spin chain Hamiltonians which correspond to perturbative matrix models and 
gauge theories have a somewhat restricted form with respect to the general
construction in the previous section. 
This is because the matrix model Hamiltonian and the gauge theory
dilatation generator receive perturbative corrections which 
can be computed from Feynman diagrams.
We illustrate the below arguments in \figref{fig:diaginter}.
First of all, it is straightforward to see that a connected planar Feynman diagram
at order $\lambda^{k}$ can attach to $k+1$ nearest neighbors only:
The diagram contains $j$ four-valent and $2k-2j$ three-valent vertices
and therefore has no more than $2k+2$ external legs. These are
split into ingoing legs, which connect to the state,
and outgoing legs, which become fields of the new state.
Here we focus on sectors where the number of spins in the state is preserved,
hence an interaction at $\order{\lambda^k}$ acts on $k+1$ adjacent spin sites
at a time. This is precisely the same constraint as for $\lambda \,\Q_2(\lambda)$,
see \secref{sec:spin-chain}.
Note that the additional power of $\lambda$ was introduced because the
interactions at $\order{\lambda^0}$ are trivial: they do not contain vertices. 
A contribution at $\order{\lambda^{k+1}}$ to the planar
gauge theory dilatation operator $\dil$ or the planar matrix model Hamiltonian $\ham$
is equivalent to a $\order{\lambda^k}$ contribution in the spin chain Hamiltonian $\Q_2$.
Explicitly, the spin chain Hamiltonian is defined, respectively, as
\[\label{eq:ModelHam}
  \dil(\lambda) = \dil_0 + \lambda \Q_2(\lambda)\qquad\mbox{or}\qquad
  \ham(\lambda) = \ham_0 + \lambda \Q_2(\lambda) \; .
\]

Furthermore, we focus on matrix models or gauge theories with
at most four-valent elementary vertices. 
This has a subtle effect on the set of allowed interactions
with maximum range, c.f.~\figref{fig:diaginter}:
Any interaction should be thought of as composed from the 
elementary vertices. If there are conserved flavors, then 
the charges can only flow between the spin chain sites 
at the vertices. However, in the planar limit, a vertex
can at most permute the flavors of two nearest neighbors. 
Therefore a contribution at order $\lambda^k$ in the matrix
model can be composed from no more than $k$ nearest-neighbor permutations. 
This means that terms at $\order{\lambda^k}$ in $\Q_2$ should be built 
out of symbols $\PTerm{a_1,\ldots,a_l}$ with no more than $k+1$ entries.
In other words, the symbols in grey font in line $R=2+k$ 
of \tabref{tab:permutation-operators} are not permitted in $\Q_2^{(k)}$.
Note however, that this argument is rigorous only for
interactions with maximum range at each order. 
Interactions of subleading range may contain
internal loops, which could, in principle, 
induce arbitrary flavor flow within the interaction.
Nevertheless, the restriction appears to apply
also for interactions with non-maximal range.

The first term for which the latter restriction applies
is $\PTerm{0,1,0}$ at $\order{\lambda}$. This term
has maximum allowed range $R=3$ for $\Q_2^{(1)}$.
It however consists of three elementary permutations, 
which is incommensurate with four-valent matrix models.
This is somewhat disappointing, as this terms does
appear in $\Q_2$ in \tabref{tab:Q2}
for almost all values of the parameters.
This would seem to imply that a Hamiltonian from a
four-valent matrix model \emph{cannot be integrable}.

One immediate way of avoiding this no-go theorem is to set
$\alpha_0(\lambda)=\order{\lambda^2}$ instead
of $\alpha_0(\lambda)=\order{\lambda}$ as 
it would be allowed in general.
Then the term $\PTerm{0,1,0}$ would arise only at $\order{\lambda^2}$ where
it is indeed permitted. Unfortunately, this makes the contribution at
$\order{\lambda}$ trivial which is not what is expected.

The only suitable evasion of the no-go theorem is to restrict
to $\alg{gl}(2)$ symmetry. Then there is an epsilon identity 
which expresses $\PTerm{0,1,0}$ as a linear combination of
$\PTerm{0,1},\PTerm{1,0},\PTerm{0},\PTerm{}$,
see \eqref{eqn:permutations-su2-rule},
all of which are allowed.
Remarkably, this works even at higher orders, we merely
need to postpone the appearance of $\beta_{r,s}(\lambda)$
to
\[
\beta_{r,s}(\lambda)=\order{\lambda^{r+s-2}} 
\]
instead of $\beta_{r,s}(\lambda)=\order{\lambda^{s-1}}$.
The reason is that the structure coupling to $\beta_{r,s}$ 
appears to be composed from $r+s-1$ elementary permutations.

Finally, the matrix models usually preserve a $\grp{U}(N)$ 
charge conjugation symmetry. This corresponds to parity conjugation 
for spin chains and hence we need to set parity-violating parameters to zero
\<
\alpha_{\ell}(\lambda)\eq 0\qquad\mbox{for } \ell\mbox{ odd},
\nln
\beta_{r,s}(\lambda)\eq 0\qquad\mbox{for } r+s\mbox{ even},
\nln
\gamma_{r,s}(\lambda)\eq 0 \qquad\mbox{for } r+s\mbox{ odd}.
\>
Parity conjugation sends all $u_{\ell,k}$ to $-u_{\ell,k}$.
It is then easy to see that the Bethe equations 
\eqref{eq:BetheMain,eq:BetheAux,eq:BetheAuxEnd}
remain valid and the charges 
\eqref{eq:chargeeigen}
are not modified if the above parameters 
are removed.

\section{Coefficients for specific models} 
\label{sec:coeffs}

Here we list the coefficients for some specific models 
studied in the literature. The Bethe ansatz for the $\alg{su}(2)$ sector 
of $\superN=4$ superconformal Yang-Mill theory \cite{Beisert:2003tq}
was found in \cite{Serban:2004jf,Beisert:2004hm}. 
Setting $\lambda = \gym^2N/16\pi^2$, we have
\[\label{eq:BDS}
\alpha_\ell(\lambda)  = \tilde\alpha_\ell(\lambda) = \lambda\, \delta_{\ell,0} \; , \qquad
\beta_{r,s}(\lambda)  = 0 \; , \qquad
\gamma_{2,r}(\lambda) = 2\,\delta_{2,r} \; .
\]
This result is rigorous at $\order{\lambda^2}$
corresponding to three loops in gauge theory~\cite{Beisert:2003ys,Eden:2004ua}.
Beyond that, there may or may not arise corrections. 

A corresponding Bethe ansatz for the $\alg{su}(2)$ sector of 
quantum strings on $AdS_5\times S^5$ 
was conjectured in \cite{Arutyunov:2004vx}.
Here $\lambda = R^4/16\pi^2 {\alpha'}^2$ is the AdS/CFT dual parameter and we have
\[\label{eq:AFS}
\alpha_\ell(\lambda)  = \tilde\alpha_\ell(\lambda) = \lambda\, \delta_{\ell,0} \; , \qquad
\beta_{r,s}(\lambda)  = \lambda^r \delta_{r+1,s} \; , \qquad
\gamma_{2,r}(\lambda) = 2\,\delta_{2,r} \; .
\]
The coefficients $\beta_{r,s}(\lambda)$ are known to 
receive corrections at subleading orders in $1/\sqrt{\lambda}$
\cite{Beisert:2005cw}. 
Note that this Bethe ansatz is not based on a spin chain.
Nevertheless, the structures in this Bethe ansatz are compatible
with the form we have derived here. 
Thus, when we extrapolate \eqref{eq:AFS} naively to small $\lambda$
we do obtain a spin chain as observed in \cite{Beisert:2004jw}.
The Hamiltonian of this `string' chain agrees with the one given in \tabref{tab:Q2}.

For the $\alg{su}(2)$ sector of plane-wave matrix theory, see \secref{sec:pwmt}, the Hamiltonian was
computed up to fourth order in $\lambda = 2N/M^3$ \cite{Fischbacher:2004iu}.
The non-zero coefficients of the corresponding spin chain system are
\begin{align}
  \alpha_0(\lambda)     & = \tilde\alpha_0(\lambda) = \lambda - \tfrac{7}{2} \lambda^2 + \tfrac{71}{2} \lambda^3 + \order{\lambda^4} \; , \nln
  \beta_{2,3}(\lambda)  & = \tfrac{13}{16} \lambda^3 + \order{\lambda^4} \; ,\nln
  \gamma_{2,2}(\lambda) & = 2 - 7 \lambda + 71 \lambda^2 - \tfrac{7767}{8} \lambda^3 + \order{\lambda^4} \; .
\end{align}
A similar four-valent matrix theory with $\alg{so}(6)$ symmetry
has an $\alg{su}(2)$ sector whose Hamiltonian is integrable only up to order $\order{\lambda^3}$
\cite{Klose:2005cv}. The coefficients to describe the spectrum at this order are given by
\begin{align}
  \alpha_0(\lambda)     & = \tilde\alpha_0(\lambda) = \lambda - \tfrac{55}{2} \lambda^2 \; , \nln
  \beta_{r,s}(\lambda)  & = 0 \; , \nln
  \gamma_{2,0}(\lambda) & = 9 - \tfrac{615}{4} \lambda + \tfrac{39123}{8} \lambda^2 \; , \nln
  \gamma_{2,2}(\lambda) & = 2 - 37 \lambda + \tfrac{4601}{4} \lambda^2 \; .
\end{align}

Finally, let us consider the hyperbolic Inozemtsev spin chain \cite{Inozemtsev:1989yq,Inozemtsev:2002vb}. 
This model is characterized by a $\Q_2$ 
which consists only of permutations of two (not necessarily adjacent) spins. 
The interaction strength is governed by the Weierstrass function 
with imaginary period $i\pi/\kappa$. 
Up to $\order{\lambda^3}$, where $\lambda = \sum_{n=1}^\infty 4\pi^2/\sinh^2(\kappa n)$, 
our coefficients need to be adjusted as follows \cite{Beisert:2004hm}:
\begin{align}
  \alpha_0(\lambda)          & = \tilde\alpha_0(\lambda) = \lambda+\order{\lambda^4} \; , \nln
  \alpha_2(\lambda)          & = \tilde\alpha_2(\lambda) = \lambda^3+\order{\lambda^4} \; , \nln
  \beta_{r,s}(\lambda)       & = 0 \; , \nln
  \gamma_{2,2}(\lambda)      & = 2 + 6 \lambda - 20 \lambda^2 + 120 \lambda^3+\order{\lambda^4} \; , \nln
  \gamma_{2,4}(\lambda)      & = 6 \lambda^2 - 30 \lambda^3+\order{\lambda^4} \; .
\end{align}

\section{Plane-wave matrix theory not fully integrable} 
\label{sec:pwmt}

In this section we show that PWMT is not completely integrable
because it possesses a non-integrable $\alg{su}(3)$ subsector. 
We give two different arguments to prove this statement. 
For the first reasoning we find all closed subsectors of 
PWMT among which there is a $\alg{su}(3)$ subsector 
with spins in the fundamental representation. 
As an immediate consequence of the discussion in \secref{sec:gauge-theory}, 
this sector cannot be integrable since the 
PWMT interactions are indeed at most four-valent. 
The second reasoning is based on the parity invariance of PWMT 
and demonstrates that there does not exist any parity-odd commuting charge 
in the considered $\alg{su}(3)$ subsector and hence no integrability.

Let us recall some basic properties of PWMT. 
Its global symmetry algebra is $\alg{su}(4|2)$. 
The elementary matrix excitations $\Md_{AB}$ transform 
under the seventeen-dimensional 
anti-symmetric two-tensor representation of this algebra. 
The fundamental $\alg{su}(4|2)$ index $A = 1,2,3,4\mathpunct{|}5,6 = (a|\alpha)$ 
splits into a commuting $\alg{su}(4)$ index $a$ 
and an anti-commuting $\alg{su}(2)$ index $\alpha$. 
A generic single-trace state is given by
\[
  \Tr \bigbrk{ \Md_{A_1 B_1} \Md_{A_2 B_2} \ldots \Md_{A_L B_L} } \state{0} \; .
\]
It corresponds to a cyclic state of an $\alg{su}(4|2)$ spin chain 
of length $L$ with spins transforming in the representation
with Dynkin labels $\dynkin{0,1,0\mathpunct{|}0\mathpunct{|}0}$.

A subsector of the theory is spanned by the states 
which are built from only a subset of all elementary excitations. 
A subsector is called closed, if the states of this subsector 
do not mix with other states under the application of the Hamiltonian. 
E.g.~the closed subsector $\alg{su}(2)$ discussed in \secref{sec:spin-chain} 
is obtained from the two bosonic excitations
\[ \label{eq:su2sector}
  \{ \Zd \equiv \Md_{12} \; , \; \Wd \equiv \Md_{13} \} \; .
\]
In order to find a closed subsector, 
we need to look for conserved `quantum numbers' 
that can be used to define the subsector. 
These quantum numbers are provided by 
the eigenvalues of the Cartan generators of $\alg{su}(4|2)$:
\[
  \A_1 = n_1 - n_2
  \; , \;
  \A_2 = n_2 - n_3
  \; , \;
  \A_3 = n_3 - n_4
  \; , \;
  \A_4 = n_4 + n_5
  \; , \;
  \A_5 = n_5 - n_6
  \; ,
\]
where $n_i$ is the number of indices in a state 
which assume the value $i$. 
These are appropriate quantum numbers 
because the free PWMT Hamiltonian 
is actually an element of the Cartan subalgebra 
and thus preserves these values.

Now, any linear combination $\A$ of these Cartan generators 
which is a positive semi-definite operator 
on the elementary excitations
defines a closed subsector \cite{Beisert:2003jj,Beisert:2004ry}. 
The excitations of this subsector are the ones which lie in the kernel of $\A$.

As an example we may take $\A = 2 \A_4 - \A_5 = 2 n_4 + n_5 + n_6$. 
Then we have $\A = 0$ for
\[ \label{eq:su3sector}
  \{ \Zd \equiv \Md_{12} \; , \; \Wd \equiv \Md_{13} \; , \; \bYd \equiv \Md_{23} \} \; .
\]
and $\A > 0$ for all other excitations. 
Hence any state with $\A=0$ can only be built 
from these three special oscillators, 
as any other excitation has a strictly positive eigenvalue of $\A$. 
In fact these three oscillators transform 
according to the fundamental representation of the $\alg{su}(3)$
subalgebra which commutes with $\A$. 
In this way, $\A$ defines a closed fundamental $\alg{su}(3)$ subsector. 

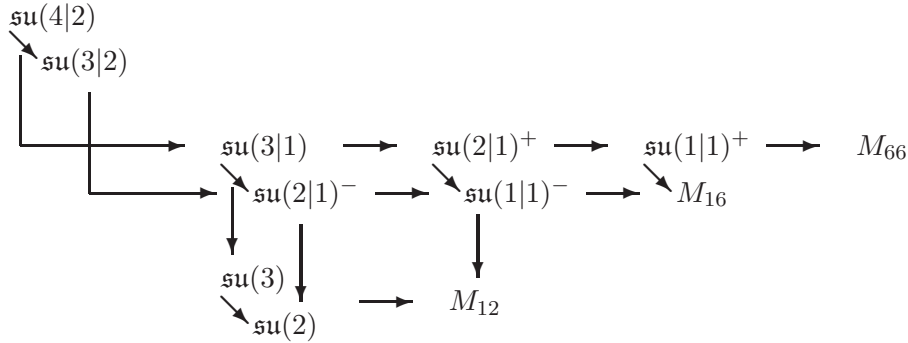
\begin{figure}\centering
\setlength{\unitlength}{1pt}%
\small\thicklines%
\begin{picture}(340,125)(0,-125)
\put(  0,   0){\makebox(0,0)[tl]{\vphantom{\big|}$\alg{su}(4|2)$}}%
\put(  0, -13){\vector(1,-1){10}}%
\put( 12, -18){\makebox(0,0)[tl]{\vphantom{\big|}$\alg{su}(3|2)$}}%
\put(  4, -22){\line(0,-1){34}}%
\put(  4, -56){\vector(1,0){62}}%
\put( 30, -36){\line(0,-1){38}}%
\put( 30, -74){\vector(1,0){48}}%
\put( 80, -50){\makebox(0,0)[tl]{\vphantom{\big|}$\alg{su}(3|1)$}}%
\put( 80, -63){\vector(1,-1){10}}%
\put( 92, -68){\makebox(0,0)[tl]{\vphantom{\big|}$\alg{su}(2|1)^-$}}%
\put(126, -56){\vector(1,0){20}}%
\put(138, -74){\vector(1,0){20}}%
\put(160, -50){\makebox(0,0)[tl]{\vphantom{\big|}$\alg{su}(2|1)^+$}}%
\put(160, -63){\vector(1,-1){10}}%
\put(172, -68){\makebox(0,0)[tl]{\vphantom{\big|}$\alg{su}(1|1)^-$}}%
\put(206, -56){\vector(1,0){20}}%
\put(218, -74){\vector(1,0){20}}%
\put(240, -50){\makebox(0,0)[tl]{\vphantom{\big|}$\alg{su}(1|1)^+$}}%
\put(240, -63){\vector(1,-1){10}}%
\put(252, -68){\makebox(0,0)[tl]{\vphantom{\big|}$M_{16}$}}%
\put(286, -56){\vector(1,0){20}}%
\put(320, -50){\makebox(0,0)[tl]{\vphantom{\big|}$M_{66}$}}%
\put( 84, -72){\vector(0,-1){25}}%
\put(110, -86){\vector(0,-1){29}}%
\put(177, -82){\vector(0,-1){24}}%
\put( 80,-100){\makebox(0,0)[tl]{\vphantom{\big|}$\alg{su}(3)$}}%
\put( 80,-113){\vector(1,-1){10}}%
\put( 92,-118){\makebox(0,0)[tl]{\vphantom{\big|}$\alg{su}(2)$}}%
\put(132,-115){\vector(1,0){20}}%
\put(166,-109){\makebox(0,0)[tl]{\vphantom{\big|}$M_{12}$}}%
\end{picture}
\caption{Structure of closed sectors of PWMT.
When moving along a vertical/horizontal line, 
the index range of $\alpha$/$a$ is reduced by one, 
i.e.~one index value is deactivated. 
When moving along a diagonal line, one index $a$ is fixed to a particular value.}
\label{fig:sectorconnect}
\end{figure}

PWMT possesses a number of closed subsectors
with fields in different representations; we list all of them in \tabref{tab:sectors}
and show their relations in \figref{fig:sectorconnect}. 
According to the general discussion in \secref{sec:gauge-theory},
it is the presence of the fundamental $\alg{su}(3)$ subsector that 
spoils the full integrability of the model beyond leading order.

\begin{table}\centering
$\begin{array}{|l|l|c|r|}\hline
\multicolumn{1}{|c|}{\mbox{sector}}&
\multicolumn{1}{c|}{\mbox{Dynkin labels}}&
\multicolumn{1}{c|}{\mbox{excitations}}&
\multicolumn{1}{c|}{\A}
\\\hline
\mbox{light boson $\Md_{12}$}&                                            & {1|0|0} & -\A_1\phantom{\mathord{}+\A_2}      +2\A_3+5\A_4-2\A_5 \\
\mbox{fermion $\Md_{16}$}    &                                            & {0|1|0} & -\A_1+ \A_2+3\A_3+5\A_4-\phantom{2}\A_5 \\
\mbox{heavy boson $\Md_{66}$}&                                            & {0|0|1} &  \A_1+2\A_2+3\A_3+4\A_4\phantom{\mathord{}-2\A_5}       \\ 
\alg{su}(2)                  & \dynkin{1}                                 & {2|0|0} & -\A_1\phantom{\mathord{}+\A_2}      +\phantom{2}\A_3+4\A_4-2\A_5 \\
\alg{su}(1|1)^- \mbox{ (light)} & \dynkin{1}                                 & {1|1|0} & -\A_1\phantom{\mathord{}+\A_2}      +2\A_3+4\A_4-\phantom{2}\A_5 \\
\alg{su}(1|1)^+ \mbox{ (heavy)} & \dynkin{2}                                 & {0|1|1} &        \A_2+2\A_3+3\A_4\phantom{\mathord{}-2\A_5}       \\
\alg{su}(2|1)^- \mbox{ (light)} & \dynkin{1\mathpunct{|}0}                   & {2|1|0} & -\A_1\phantom{\mathord{}+\A_2}+\phantom{2} \A_3+3\A_4-\phantom{2}\A_5 \\
\alg{su}(2|1)^+ \mbox{ (heavy)} & \dynkin{0\mathpunct{|}1}                   & {1|2|1} &              \A_3+2\A_4\phantom{\mathord{}-2\A_5}       \\
\alg{su}(3)                  & \dynkin{0,1}                               & {3|0|0} &                   2\A_4-\phantom{2}\A_5 \\
\alg{su}(3|1)                & \dynkin{0,1\mathpunct{|}0}                 & {3|3|1} &                    \A_4\phantom{\mathord{}-2\A_5}       \\
\alg{su}(3|2)                & \dynkin{1,0\mathpunct{|}0\mathpunct{|}0}   & {3|2|0} & -\A_1\phantom{\mathord{}+\A_2}+\phantom{2} \A_3+2\A_4-\phantom{2}\A_5 \\
\alg{su}(4|2)                & \dynkin{0,1,0\mathpunct{|}0\mathpunct{|}0} & {6|8|3} & 0                             \\\hline
\end{array}$

\caption{Closed sectors of PWMT with residual symmetry algebra, representation of spins, 
excitation content in terms of light bosons, fermions and heavy bosons, 
and a choice for the Cartan generator $\A$ whose kernel describes the sector.}
\label{tab:sectors}
\end{table}

Now, we also verify the non-integrability of this sector by an explicit calculation. 
As the PWMT Hamiltonian is parity invariant, 
we may work with states of definite parity $P=\pm$. 
The parity-odd charges then have the property to pair up multiplets 
of opposite parity 
(or annihilate the multiplets 
in case there is no suitable partner multiplet 
of opposite parity in the spectrum). 
The commutation of these charges with the Hamiltonian 
implies a degenerate energy eigenvalue for both states of a parity pair. 
If this degeneracy is violated in a parity conserving theory, 
integrability must be broken \cite{Grabowski:1995rb,Beisert:2003tq}.

So let us consider multiplets 
with Dynkin labels $\dynkin{1,1,1\mathpunct{|}1\mathpunct{|}0}$ of $\alg{su}(4|2)$. 
This multiplet has representatives in the $\alg{su}(3)$ subsector \eqref{eq:su3sector}.
These are not at the same time contained in the $\alg{su}(2)$ subsector \eqref{eq:su2sector}
where degeneracy would be guaranteed by the
established integrability at the considered perturbative order 
\cite{Klose:2003qc}.
This multiplet is realized twice in the model,
once with positive parity and $\alg{su}(3)$ highest weight state
\[
\begin{split}
\state{+} = \ & \bigl( \Tr \bYd \acomm{\comm{\Wd }{\Zd }}{\Zd} \comm{\Wd}{\Zd} \\
              & +      \Tr \Wd  \acomm{\comm{\Zd }{\bYd}}{\Zd} \comm{\Wd}{\Zd} \\
              & +      \Tr \Zd  \acomm{\comm{\bYd}{\Wd }}{\Zd} \comm{\Wd}{\Zd} \bigr) \state{0}
\end{split}
\]
and another time with negative parity and highest weight state
\[
\begin{split}
\state{-} = \ & \bigl(
  - \Tr \Zd \Zd \Wd \Wd \Zd \bYd
  - \Tr \Zd \Zd \Wd \Zd \Wd \bYd
  + \Tr \Zd \Zd \bYd \Wd \Zd \Wd \\
& + \Tr \Zd \Zd \bYd \Zd \Wd \Wd
  + \Tr \Zd \Zd \Zd \Wd \Wd \bYd
  - \Tr \Zd \Zd \Zd \bYd \Wd \Wd \bigr) \state{0} \; .
\end{split}
\]
When we act with the spin chain Hamiltonian
\begin{align}
  \Q_2^{(0)} & =    2 \PTerm{} -  2 \PTerm{0} \; , \\
  \Q_2^{(1)} & = - 15 \PTerm{} + 19 \PTerm{0} - 2(\PTerm{0,1}+\PTerm{1,0}) \; ,
\end{align}
corresponding to PWMT in the $\alg{su}(3)$ sector at two loops \cite{Klose:2003qc},
we find that the zeroth order degeneracy is lifted by the first order corrections:
\begin{align}
(\Q_2^{(0)} + \lambda \Q_2^{(1)} ) \state{+} & = ( 10 - 73 \lambda ) \state{+} \; , \\
(\Q_2^{(0)} + \lambda \Q_2^{(1)} ) \state{-} & = ( 10 - 65 \lambda ) \state{-} \; .
\end{align} 

Recalling that PWMT is derived from SYM and therefore, in some sense, 
contained in SYM, it is natural to ask why the integrability 
is not broken in the mother theory, too. 
The reason is simply that the $\alg{su}(3)$ sector 
containing the fields \eqref{eq:su3sector} is not closed in SYM. 
There occurs mixing with two fermionic fields 
which make it an integrable $\alg{su}(3|2)$ sector \cite{Beisert:2003ys}. 
In the derivation of PWMT these fermionic degrees of freedom 
have been projected out and integrability has vanished with them.

\section{Conclusions and outlook}

We have investigated the structure of the Hamiltonian and the Bethe ansatz 
of the \emph{most general} integrable spin chain with the following properties:
\begin{enum}
\item spins transforming in the fundamental representation of $\alg{gl}(n)$ and
\item long-range interactions suppressed by $\lambda^{R-2}$ where $R$ is their range.
\end{enum}
In the current absence of rigorous proofs of integrability for these systems,
we have made the assumption that it is sufficient to ensure
the existence of at least one conserved charge,
cf.~\cite{Grabowski:1995rb}.

The derivation has been carried out explicitly 
up to fourth order in the formal expansion parameter $\lambda$,
i.e.~up to interactions of range six.
The discovered form of the Bethe ansatz equations, 
however, suggests an elegant extension to all orders.
According to this, the moduli of the system 
are expressed through four sets of analytic functions 
in the expansion parameter $\lambda$: 
\begin{enum}
\item 
the rapidity maps $\alpha_{\ell}(\lambda)$ with $\ell = 0,1,2,\ldots$ 
are of order $\order{\lambda^{\ell+1}}$ 
and describe the propagation of spins along the spin chain, 

\item 
the dressing phases $\beta_{r,s}(\lambda)$ with $r,s = 2,3,4,\ldots$\,,
$r<s$ are of order $\order{\lambda^{s-1}}$ and describe the interaction of spins, 

\item 
the elements of the normalization matrix $\gamma_{r,0}(\lambda)$, $\gamma_{r,s}(\lambda)$ 
with $r,s = 2,3,4,\ldots$ are of order $\order{\lambda^{\max(s-r,0)}}$ 
and describe the mixing between the commuting charges, and 

\item 
the similarity parameters $\epsilon_{k,\ell}(\lambda)$ 
with $\ell = 1,\ldots,(k+1)!-2k!+(k-1)!-1$ 
are of order $\order{\lambda^k}$, describe unitary changes 
of the basis and consequently have no influence on the energy spectrum 
or the Bethe equations. 
\end{enum}
Supposing that our conjecture is correct, 
every spin chain with the aforementioned properties 
is realized by one particular choice of the moduli.
Conversely, for every choice of moduli 
there would be a corresponding spin chain Hamiltonian.
We have proven the validity of our conjecture up to fourth order in $\lambda$.

We have found that, remarkably, the form of the Hamiltonian 
and the set of physical moduli%
\footnote{The number of `unphysical' parameters $\epsilon_{r,s}$
may depend on the rank because of structures
which are identically zero for small values of $n$.}
is universal; it is independent of the rank $n$ 
of the symmetry algebra $\alg{gl}(n)$. 
Another observation is that any nearest-neighbor Hamiltonian 
which is deformed by the interactions we have described, must have 
interactions of unbounded range in order to be exactly integrable
\cite{Beisert:2003tq}; it must be a long-range spin chain.

We furthermore found that the spin chain Hamiltonian derived from PWMT
in an $\alg{su}(3)$ subsector is not of the general form 
and therefore not integrable, 
a fact that we also demonstrated in an explicit calculation. 
This also implies that PWMT is \emph{not} integrable as a whole.
The established data concerning integrability in PWMT are hence: 
\begin{enum}
\item
one-loop integrability in the complete model \cite{Beisert:2003jj,Beisert:2003yb},
\item
at least three-loop integrability in the $\alg{su}(3|2)$ sector \cite{Beisert:2003ys} and
\item
at least four-loop integrability in the $\alg{su}(2)$ sector
\cite{Klose:2003qc,Fischbacher:2004iu}.
\end{enum}
Here we have added: 
\begin{enum}
\item
no integrability in the $\alg{su}(3)$ sector beyond one-loop and therefore
\item
no integrability of the full Hamiltonian beyond one-loop.
\end{enum}
Note that the perturbative non-integrability of the 
complete model does not exclude potential integrability in the 
$\alg{su}(3|2)$ and $\alg{su}(2)$ sectors to all orders.
\bigskip

To gain a better understanding of integrability 
in PWMT, it would be interesting to derive and investigate
the Hamiltonian in the various non-trivial vacua
of the theory \cite{Dasgupta:2002ru,Lin:2005nh}.
As the elementary representation content 
in vacua with increasing M5-brane number
approaches that of $\superN=4$ SYM,
there is some hope of enhancing 
or even recovering full integrability.

As a further development of our results one could consider the generalization 
to spins in an arbitrary representation of $\alg{gl}(n)$.
Do these generalizations lead to integrable systems, what are their moduli 
and what are the Bethe equations?
As a first step, one might consider totally symmetric and anti-symmetric 
representations. These have only one non-vanishing Dynkin label
and the nested Bethe ansatz is conceptually simpler 
than in the case of general representations.

\subsection*{Acknowledgments}

We are grateful to J.~Maldacena and M.~Staudacher for discussions.
We also would like to thank K.~Zarembo for comments on the manuscript.
The work of N.~B.~is supported in part by
the U.S.~National Science Foundation Grant No.~PHY02-43680. Any
opinions, findings and conclusions or recommendations expressed in
this material are those of the authors and do not necessarily
reflect the views of the National Science Foundation.
The work of T.~K.~is supported by the G\"oran Gustafsson Foundation.  

\appendix

\section{Permutation operators}
\label{sec:perms}

\begin{figure}[b]\centering
$\PTerm{2,1,0,2}
\to
\begin{minipage}{145pt}
\setlength{\unitlength}{1pt}%
\small\thicklines%
\begin{picture}(145,110)(-5,-15)
\put(  0, 0){\circle*{5}}%
\put( 15, 0){\circle*{5}}%
\put( 30, 0){\circle*{5}}%
\put( 45, 0){\circle*{5}}%
\put(  0,-5){\makebox(0,0)[t]{0}}%
\put( 15,-5){\makebox(0,0)[t]{1}}%
\put( 30,-5){\makebox(0,0)[t]{2}}%
\put( 45,-5){\makebox(0,0)[t]{3}}%
\put(110, 5){\makebox(0,0)[cl]{$\}$}}%
\put(  0, 0){\line(0,1){15}}%
\put( 15, 0){\line(0,1){15}}%
\put( 30, 0){\line(0,1){15}}%
\put( 45, 0){\line(0,1){15}}%
\put(100,20){\makebox(0,0)[cl]{$2\hateq\calP_{23}$}}%
\put(  0,15){\line(0,1){15}}%
\put( 15,15){\line(0,1){15}}%
\put( 30,15){\line(1,1){15}}%
\put( 45,15){\line(-1,1){15}}%
\put( 90,35){\makebox(0,0)[cl]{$0\hateq\calP_{01},$}}%
\put(  0,30){\line(1,1){15}}%
\put( 15,30){\line(-1,1){15}}%
\put( 30,30){\line(0,1){15}}%
\put( 45,30){\line(0,1){15}}%
\put( 80,50){\makebox(0,0)[cl]{$1\hateq\calP_{12},$}}%
\put(  0,45){\line(0,1){15}}%
\put( 15,45){\line(1,1){15}}%
\put( 30,45){\line(-1,1){15}}%
\put( 45,45){\line(0,1){15}}%
\put( 70,65){\makebox(0,0)[cl]{$2\hateq\calP_{23},$}}%
\put(  0,60){\line(0,1){15}}%
\put( 15,60){\line(0,1){15}}%
\put( 30,60){\line(1,1){15}}%
\put( 45,60){\line(-1,1){15}}%
\put( 60,80){\makebox(0,0)[cl]{$\{$}}%
\put(  0,75){\line(0,1){15}}%
\put( 15,75){\line(0,1){15}}%
\put( 30,75){\line(0,1){15}}%
\put( 45,75){\line(0,1){15}}%
\end{picture}
\end{minipage}
=
\calP_{23}
\calP_{12}
\calP_{01}
\calP_{23}$.
\caption{Graphical representation of the permutation symbol
$\PTerm{2,1,0,2}$.}
\label{fig:sample-permutation}
\end{figure}
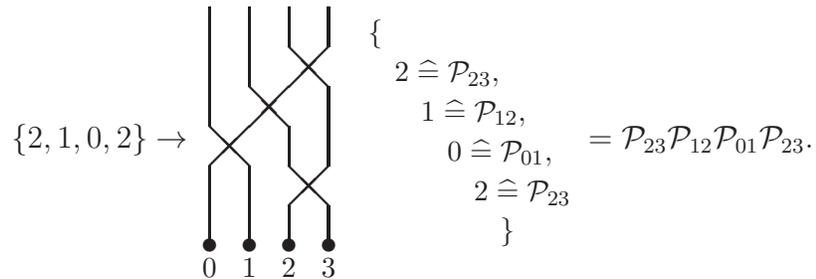

It is very convenient to represent the spin chain operator 
by a product of nearest-neighbor permutation denoted 
as in \eqref{eqn:permutations} by
\[
  \PTerm{a_1,a_2,\ldots,a_l} := \sum_{i=1}^{\calL} \calP_{i+a_1,i+a_1+1} \calP_{i+a_2,i+a_2+1} \cdots \calP_{i+a_l,i+a_l+1} \; , 
\qquad
  \PTerm{}                   := \calL \; ,
\]
where $\calL$ measures the spin chain length $L$. 
The indices are clearly to be understood modulo $L$. 
In terms of Pauli matrices, we have e.g.
\[
  \PTerm{0} = \sum_{i=1}^L \calP_{i,i+1} = \sum_{i=1}^L \frac{1}{2}(\unit + \vec\sigma_{i}\cdot\vec\sigma_{i+1}) \; .
\]
It is helpful to visualize the sequence of transpositions 
in such an operator as for example in \figref{fig:sample-permutation}.
We call the product of transpositions 
(i.e.~the right hand side of \eqref{eqn:permutations} without the sum),
a \emph{local interaction}. 
The total action of a permutation is given by applying the local interaction 
to all different positions along the state and summing these contributions. 
The number $R := \max\set{a_i} - \min\set{a_i} + 2$ 
is called the \emph{range} of the permutation $\PTerm{a_1,a_2,\ldots,a_l}$, 
since this is the number of adjacent fields 
which are hit by the local interaction at once. 
The quantity $l$ is called 
the \emph{length} of the permutation 
and counts the number of elementary transpositions. 
In the language of Feynman diagrams $l$ corresponds 
to the number of four-valent vertices as discussed in \secref{sec:gauge-theory}.

There are some relations between these permutations operators:
\begin{align}\label{eqn:permutations-trivial-rules}
  \PTerm{\mathline,a,a,\mathline} & = \PTerm{\mathline,\mathline} \; ,     \nln
  \PTerm{\mathline,a,b,\mathline} & = \PTerm{\mathline,b,a,\mathline} \; ,  \qquad\qquad\qquad \mbox{for $\abs{a-b} \ge 2$}  \nln
  \PTerm{\mathline,a,a+1,a,\mathline} & = \PTerm{\mathline,a+1,a,a+1,\mathline} \; ,  \nln
  \PTerm{a_1,a_2,\ldots}    & = \PTerm{a_1+b,a_2+b,\ldots} \; . 
\end{align}
Furthermore we can use the fact that the anti-symmetrizer 
of $n+1$ spins vanishes for $\alg{gl}(n)$ and therefore replace
\[
\PTerm{\mathline, \quad 0, \quad 1,0, \quad 2,1,0, \quad 3,2,1,0, \quad\ldots , \quad n-1, n-2,\ldots,1,0,\quad \mathline}
\]
by a sum of permutations of shorter range. E.g.~in $\alg{gl}(2)$ we have
\<\label{eqn:permutations-su2-rule}
  \PTerm{\mathline,0,1,0,\mathline}\eq \PTerm{\mathline,0,1,\mathline}
                                         + \PTerm{\mathline,1,0,\mathline} \nl
                                        - \PTerm{\mathline,0,\mathline}
                                         - \PTerm{\mathline,1,\mathline}
                                         + \PTerm{\mathline,\mathline} \; .
\>
Finally, we define 
hermitian conjugation by
\[ \label{eqn:permutation-operators-hermitian-conjugation}
  \PTerm{a_1,\ldots,a_l}^\dagger = \PTerm{a_l,\ldots,a_1} \; ,
\]
and parity conjugation by
\[ \label{eqn:permutation-operators-parity-conjugation}
  \PTerm{a_1,\ldots,a_l}^\sharp = \PTerm{-a_1,\ldots,-a_l} \; .
\]

\bibliographystyle{nb}
\bibliography{GLNlong}

\vspace{\fill}

\begin{table}[h]
\begin{align}
\bar\Q_2(\lambda) & =
              ( \PTerm{} - \PTerm{0} ) \nn\\[3mm]
           &      + \alpha_{0}(\lambda)\, ( - 3 \PTerm{} + 4 \PTerm{0} - \PTerm{0,1,0} ) \nn\\[3mm]
           & + \alpha_{0}(\lambda)^2 
               ( 20 \PTerm{} - 29 \PTerm{0} + 10 \PTerm{0,1,0} - \PTerm{0,1,2} - \PTerm{2,1,0} + \PTerm{0,2,1} + \PTerm{1,0,2} \nln 
           &   \qquad - \PTerm{0,1,2,1,0} ) \nln
           & + \ihalf \alpha_{1}(\lambda)\,
               ( - 6 \PTerm{0,1} +6 \PTerm{1,0} + \PTerm{0,1,2,1} - \PTerm{1,2,1,0}
                   + \PTerm{0,1,0,2} - \PTerm{0,2,1,0} ) \nln
           & + \tfrac{1}{2} \beta_{2,3}(\lambda)\,
               ( -4 \PTerm{} + 8 \PTerm{0} - 2 \PTerm{0,1} - 2 \PTerm{1,0} - 2 \PTerm{0,2} \nln
           &   \qquad  - 2 \PTerm{0,1,2} - 2 \PTerm{2,1,0}+ 2 \PTerm{0,2,1} + 2 \PTerm{1,0,2}  \nln
           &   \qquad + \PTerm{0,1,2,1} + \PTerm{1,2,1,0} + \PTerm{0,1,0,2} + \PTerm{0,2,1,0}- 2 \PTerm{1,0,2,1} ) \nln
           & + i\epsilon_{2,1}(\lambda)\, ( \PTerm{1,0,2} - \PTerm{0,2,1} ) \nln
           & + i\epsilon_{2,2}(\lambda)\, (-\PTerm{0,1,2,1} + \PTerm{1,2,1,0}
                   + \PTerm{0,1,0,2} - \PTerm{0,2,1,0} )\nn\\[3mm]
           & +\order{\lambda^3}\nn
\end{align}
\caption{Normalized Hamiltonian printed up to second order}
\label{tab:Q2}
\end{table}

\vspace*{\fill}

\begin{table}
\begin{align}
\bar \Q_3(\lambda) & = \tfrac{i}{2} ( \PTerm{0,1} - \PTerm{1,0} ) \nn\\[3mm]
              & + \tfrac{i}{2} \alpha_{0}(\lambda)\,
                 ( 6 \PTerm{1,0} - 6 \PTerm{0,1} + \PTerm{0,1,2,1} - \PTerm{1,2,1,0} + \PTerm{0,1,0,2} - \PTerm{0,2,1,0} ) \nln[3mm]
           & + \tfrac{i}{2} \alpha_{0}(\lambda)^2  
               ( 46 \PTerm{0,1} - 46 \PTerm{1,0} - 12 \PTerm{0,1,0,2} - 12 \PTerm{0,1,2,1} + 2 \PTerm{0,1,2,3} \nln
           &   \qquad - 2 \PTerm{0,1,3,2} + 12 \PTerm{0,2,1,0} - 2 \PTerm{0,2,1,3} + 2 \PTerm{0,3,2,1} - 2 \PTerm{1,0,2,3} \nln
           &   \qquad + 2 \PTerm{1,0,3,2} + 12 \PTerm{1,2,1,0} + 2 \PTerm{2,1,0,3} - 2 \PTerm{3,2,1,0} + \PTerm{0,1,0,2,3,2} \nln
           &   \qquad +   \PTerm{0,1,2,1,0,3} + \PTerm{0,1,2,3,2,1} - \PTerm{0,1,3,2,1,0} - \PTerm{0,2,3,2,1,0} \nln
           &   \qquad -   \PTerm{1,2,3,2,1,0} ) \nln
           & + \tfrac{1}{4} \alpha_{1}(\lambda)\,
               (-20 \PTerm{} + 24 \PTerm{0} - 8 \PTerm{0, 1, 2} + 6 \PTerm{0, 2, 1} + 6\PTerm{1, 0, 2} - 8 \PTerm{2, 1, 0} \nln
           &   \qquad + 2 \PTerm{0, 1, 0, 2, 1} + \PTerm{0, 1, 0, 2, 3} -  \PTerm{0, 1, 0, 3, 2} - 4 \PTerm{0, 1, 2, 1, 0} \nln
           &   \qquad + 2 \PTerm{0, 1, 2, 1, 3} + \PTerm{0, 1, 2, 3, 2} - 2 \PTerm{0, 1, 3, 2, 1} - \PTerm{0, 2, 1, 0, 3} \nln
           &   \qquad - \PTerm{0, 2, 3, 2, 1} + \PTerm{0, 3, 2, 1, 0} + 2 \PTerm{1, 0, 2, 1, 0} - \PTerm{1, 0, 2, 3, 2} \nln
           &   \qquad - 2 \PTerm{1, 2, 1, 0, 3} + 2 \PTerm{1, 3, 2, 1, 0} +\PTerm{2, 3, 2, 1, 0}  )\nln
           & + \tfrac{i}{4} \beta_{2,3}(\lambda)\,
               ( - 4 \PTerm{0,1} + 4 \PTerm{1,0} + 4 \PTerm{0,1,2} + 2 \PTerm{0,1,3} + 2 \PTerm{0,2,3} - 2 \PTerm{0,3,2} \nln
           &   \qquad - 2 \PTerm{1,0,3} - 4 \PTerm{2,1,0} - 2 \PTerm{0,1,0,2} - 2 \PTerm{0,1,2,1} + 4 \PTerm{0,1,2,3} \nln
           &   \qquad - 4 \PTerm{0,1,3,2} + 2 \PTerm{0,2,1,0} - 4 \PTerm{0,2,1,3} + 4 \PTerm{0,3,2,1} - 4 \PTerm{1,0,2,3} \nln
           &   \qquad + 4 \PTerm{1,0,3,2} + 2 \PTerm{1,2,1,0} + 4 \PTerm{2,1,0,3} - 4 \PTerm{3,2,1,0} + 2 \PTerm{0,1,0,2,1} \nln
           &   \qquad -   \PTerm{0,1,0,2,3} + \PTerm{0,1,0,3,2} - 2 \PTerm{0,1,2,1,3} - \PTerm{0,1,2,3,2} - \PTerm{0,2,1,0,3} \nln
           &   \qquad + 2 \PTerm{0,2,1,3,2} - \PTerm{0,2,3,2,1} + \PTerm{0,3,2,1,0} - 2 \PTerm{1,0,2,1,0}  \nln
           &   \qquad + 2 \PTerm{1,0,2,1,3}+   \PTerm{1,0,2,3,2} - 2 \PTerm{1,0,3,2,1} + 2 \PTerm{1,3,2,1,0}  \nln
           &   \qquad - 2 \PTerm{2,1,0,3,2}+ \PTerm{2,3,2,1,0} ) \nln
           & + \tfrac{1}{2} \epsilon_{2,1}(\lambda)\,
               ( \PTerm{0, 1, 0, 2} - \PTerm{0, 1, 2, 1} - \PTerm{0, 1, 3, 2} + \PTerm{0, 2, 1, 0} \nln
           &   \qquad     +   \PTerm{0, 3, 2, 1} + \PTerm{1, 0, 2, 3} - \PTerm{1, 2, 1, 0} - \PTerm{2, 1, 0, 3} ) \nln
           & + \tfrac{1}{2} \epsilon_{2,2}(\lambda)\,
               ( \PTerm{0, 1, 0, 2, 3} - \PTerm{0, 1, 0, 3, 2} -   \PTerm{0, 1, 2, 3, 2} - \PTerm{0, 2, 1, 0, 3} \nln
           &   \qquad +   \PTerm{0, 2, 3, 2, 1} + \PTerm{0, 3, 2, 1, 0} +   \PTerm{1, 0, 2, 3, 2} - \PTerm{2, 3, 2, 1, 0} ) \nn\\[3mm]
           & + \order{\lambda^3}\nn
\end{align}
\caption{Normalized third charge printed up to second order}
\label{tab:Q3}
\end{table}

\end{document}